\documentclass[12pt,showpacs,prc]{revtex4}
\usepackage{epsfig,amssymb,amsfonts}

\makeatletter
\renewcommand{\@makefntext}[1]{\parindent=1em\noindent\hbox to 1.8em{\hss$^{\@thefnmark}$}#1}
\renewcommand{\@footnotemark}{\hbox{\mathsurround=0pt$^{\@thefnmark}$}}
\newcommand{\ftnote}[2]{\footnotemark[#1]\footnotetext[#1]{#2}}
\makeatother

\begin{document}
\title{\bf Collision damping in the $\pi ^3He\to d'N$ reaction near the
threshold}
\author{A. V. Nefediev}
\affiliation{Grupo Te\'orico de Altas Energias, Centro de F\'\i sica das Interac\c c\~oes 
Fundamentais,
Departamento de F\'\i sica, Instituto Superior T\'ecnico, Av. Rovisco Pais, P-1049-001 Lisboa, Portugal}
\affiliation{Institute of Theoretical and Experimental Physics, 117218,\\ B.Cheremushkinskaya 25, 
Moscow, Russia}
\author{M. G. Schepkin}
\affiliation{Institute of Theoretical and Experimental Physics, 117218,\\ B.Cheremushkinskaya 25, 
Moscow, Russia}
\author{H. A. Clement}
\affiliation{Physikalisches Institut, Universit{\" a}t T{\" u}bingen, Auf der
Morgenstelle 14, 72076, T{\" u}bingen, Germany}
\newcommand{\be}{\begin{equation}}
\newcommand{\ee}{\end{equation}}
\newcommand{\ds}{\displaystyle}
\newcommand{\MM}{{\cal M}}
\begin{abstract}
We present a simple quantum mechanical model exploiting the optical potential
approach for the description of collision damping
in the reaction $\pi^3He\to d'N$ near the threshold, which recently 
has been measured at TRIUMF.
The influence of the open $d'N\to NNN$ channel is taken into account. 
It leads to a suppression factor of about ten in the $d'$ survival
probability. Applications of the method to other reactions are outlined.
\end{abstract}
\maketitle

\section{Introduction}

In this paper we consider inelastic final state interaction (FSI) between 
the neutron and the hypothetical $d'$ resonance with the quantum numbers 
$T=0$, $J^P=0^-$ produced near threshold in the reaction
$\pi^3He\to d'N$. This resonance has been suggested to explain the observed
peculiar resonance-like behaviour \cite{GKD} of the pionic double charge exchange (DCX) 
on nuclei, $\pi^+ + A \to A' + \pi^-$, ranging from $^7Li$ to $^{93}Nb$ at
$T_\pi \approx 50 MeV$ (see \cite{BCS,KF,JP} and references therein).
Such a $NN$-decoupled dibaryon had been predicted by QCD-inspired models 
\cite{MADS,KMS}. In a sequence of recent papers this behaviour for the 
DCX reaction, on $Ca$ isotopes $^{42,44,48}Ca$, was qualitatively reproduced
by Nuseirat {\it et al.} \cite{N} and, for the medium and heavy nuclei,
in the generalized seniority model, by Gibbs and Wu \cite{GW}.
Still there is no conventional
model, which could demonstrate to reproduce these low-energy 
phenomena consistently for all nuclei. An independent test of the idea would be 
investigation of the  $pp\pi^-$ invariant mass spectrum in the double pion 
production, $pp \to pp\pi^-\pi^+$, since the quantum numbers of the resonance forbid 
strong coupling to $np$. Recent measurements at CELSIUS \cite{CEL} do not 
show an evidence of the resonance at the predicted level in contrast to data 
from ITEP taken at a higher proton energy \cite{V}. Thus the
situation appears not yet fully settled. Another way of searching 
for the $d'$ resonance is DCX on Helium isotopes, where $d'$ should manifest 
itself as a threshold phenomenon. At energies around $100MeV$ the total cross 
section of the reaction $^4He(\pi^+,\pi^-)$ shows an excess over conventional 
calculations which could be ascribed to the contribution of the $d'$ resonance 
\cite{JG4}. At the same time, the contribution of this resonance for both, 
$^4He(\pi^+,\pi^-)$ \cite{JG4} and $^3He(\pi^-,\pi^+)$ \cite{JG3}, occurred 
almost an order of magnitude smaller than expected. The reason for 
the suppression of the production cross sections near threshold can be due to the 
collision damping  $d'N\to NNN$, and in  the present paper we study this hypothesis using
a simple quantum mechanical model to describe the propagation of an unstable resonance in
the nuclear medium. Although singularities near production threshold always attracted
much attention and are described in many textbooks (see, {\it e.g.}, \cite{LL,BZP} and
references
therein), we are not aware of similar approaches to the final state interaction
suggested before.

The paper is organized as follows. In the second section  we present a simple quantum 
mechanical
model exploiting the optical potential approach to account for the effect of 
collision damping. In the third section we estimate the
spreading width of the $d'$ in the nuclear medium, namely, the virtual pion
exchange contribution to the spreading is calculated in detail. In the next, fourth,
section we
give our numerical estimates of collision damping and of the $d'$ survival
probability in the nuclear medium, and compare the latter to the experimental
data. The fifth section contains our conclusions and outlook.

\section{A simple quantum mechanical model}

In this section we present a simple quantum mechanical model to describe the propagation
of a resonance in the nuclear medium. As an example, we consider the behaviour 
of the hypothetical resonance $d'$ which can contribute to the pionic DCX on $^3He$.

Due to a very short interaction time of the pion with the Helium nucleus it is 
the sudden influence to be the appropriate mechanism for considering the 
$\pi^3He\to d'N$ reaction \cite{Migdal}. Thus the $d'N$ system is created at the initial
moment, $t=0$, with the wave function of Helium $\psi_{He}$ and then separates
with time as
\be
\psi_{d'N}({\bf r},t)=\psi({\bf r}-{\bf v}t,0)=\psi_{He}({\bf r}-{\bf v}t),
\label{psihe}
\ee
with the velocity $v$ defined by the energy excess over the threshold $\varepsilon$,
$$
v=\sqrt{\frac{2\varepsilon}{\mu}},
$$
where $\mu\approx\frac23m_N$ is the reduced mass of the $d'N$ system.

If the potential in the effective Schr\"odinger equation for the wave function 
(\ref{psihe}) contains an imaginary part,
\be
U({\bf r})=U_0({\bf r})+iU_1({\bf r}),\quad U_1({\bf r})<0,
\label{U}
\ee
then the probability to find the system in its initial state, $d'N$, 
decreases with time approaching a finite limit with the survival probability
being simply
\be
w_{surv}=\frac{\left|\psi_{d'N}(t\to\infty)\right|^2}{\left|\psi_{d'N}(t\to\infty)\right|^2_{U_1=0}}.
\label{surv}
\ee

Let us, for simplicity, consider the step-like form of the potential 
$U_1({\bf r})$,
\be
U_1({\bf r})=\left\{
\begin{array}{ll}
U_1={\rm const},&|{\bf r}|\leq R_1\\
0,&|{\bf r}|> R_1,
\end{array}
\right.
\label{UU1}
\ee
where the potential strength $U_1$ can be related to the elastic s-wave $d'N$
zero-angle amplitude using Born approximation:
\be
{\rm Im}F(0)\approx -\frac{\mu}{2\pi}\int U_1({\bf r}) d^3r=-\frac23\mu R_1^3U_1,
\label{Im}
\ee
whereas, according to the optical theorem, 
\be
{\rm Im}F(0)=\frac{\mu v}{4\pi}\sigma_{tot}\approx\frac{\mu v}{4\pi}\sigma_{in}.
\label{opth}
\ee
The problem of separation of the elastic and inelastic parts of the cross section
in such kind of reactions is a very subtle question. One of the relevant 
worries concerns the possible broadening of the ground state, so that a slight
change in the resonance position could have been erroneously 
arcsibed to the inelastic 
part. Bearing in mind that, in general case, 
such an effect could lead to an overestimate of
the damping in Eq.~(\ref{opth}), we still do not expect it to change 
our estimates dramatically. Indeed, from {\it a posteriori } estimate 
(see Eq.~(\ref{ratio}) below) we find the elastic part of the cross section 
to be an order of magnitude smaller than the corresponding inelastic part,
and thus it appears beyond the accuracy of the present paper.

From Eqs. (\ref{Im}), (\ref{opth}) one finds
\be
U_1=-\frac{3}{8\pi}\frac{\sigma_{in}v}{R_1^3}.
\label{U1}
\ee

Notice that we do not meet any constraints on the wave length of the resonance since, in
Born approximation, the zero-angle amplitude (\ref{Im}) does not depend on the energy of
the $d'$.

As a direct consequence of the Schr\"odinger equation one has the probability
conservation law at any moment of time in the form:
\be
\frac{d}{dt}\int_V \rho({\bf r},t)d^3r +\int_S {\bf j}({\bf r},t)d{\bf s}=
2\int_V U_1({\bf r})\rho({\bf r},t)d^3r,
\label{cons}
\ee
where
\be
\rho({\bf r},t)=|\psi_{d'N}({\bf r},t)|^2,\quad 
{\bf j}({\bf r},t)=\frac{i}{2\mu}\psi_{d'N}({\bf r},t)\nabla \psi_{d'N}^*({\bf r},t)+c.c.
\label{j}
\ee
are the probability density and the probability flux through 
the corresponding surface $S$, respectively.

The term on the r.h.s. of Eq. (\ref{cons}) describes absorption of 
particles.
In the absence of this term the density of particles would 
evolve in space and time, according to Eq. (\ref{psihe}), as
\be
\rho_0({\bf r},t)=|\psi_{d'N}({\bf r}-{\bf v}t,0)|^2=|\psi_{He}({\bf r}-{\bf v}t)|^2=
\rho_0({\bf r}-{\bf v}t,0),
\label{rho0}
\ee
if quantum spreading of the initial wave packet is neglected.

Eq. (\ref{cons}) becomes especially simple if the size of the wave packet,
hereinafter called $R_2$, is small compared to the radius of the optical potential, $R_1$.
Indeed, one can split the entire evolution time into two periods, $0\leq t\leq T$ and $t>T$, where
$T=R_1/v$ is the moment of time when $d'$ and $N$ leave the region of 
interaction. Then the solution to Eq. (\ref{cons}) reads
\be
\rho({\bf r},t)=\left\{
\begin{array}{ll}
\rho_0({\bf r},t)e^{-\Gamma t},&t\leq T\\
\rho_0({\bf r},t)e^{-\Gamma T},&t>T,
\end{array}
\right.
\label{rho}
\ee
with 
\be
\Gamma=2|U_1|.
\label{twos}
\ee
Hence the survival probability is
\be
w_{surv}=\exp\left(\frac{2U_1R_1}{v}\right)=
\exp\left(-\frac{3\sigma_{in}}{4\pi R_1^2}\right),
\label{sur0}
\ee
where we also used the relation (\ref{U1}).

The opposite limit, $R_2\ll R_1$, leads to the expression similar to (\ref{sur0})
with $R_1$ interchanged with $R_2$:
\be
w_{surv}=\exp\left(-\frac{3\sigma_{in}}{4\pi R_2^2}\right).
\label{sur1}
\ee

To see this let us consider the following anzatz for the probability density:
\be
\rho({\bf r},t)=\rho_0({\bf r},t)w(t),
\label{rho1}
\ee
so that the survival probability defined above is just $w_{surv}=w(\infty)$. 
After integration over an infinitely large volume $V$ 
(bounded by an infinitely remote surface $S$) in Eq. (\ref{cons}) one finds the following equation for the
function $w(t)$: 
\be
\frac{d}{dt}w(t)=-w(t)\int_V 2|U_1({\bf r})|\rho_0({\bf r}-{\bf v}t,0)d^3r,\quad
w(0)=1,
\label{wdot}
\ee
which can be easily solved with the result
\be
w(t)=\exp\left(-\int_0^tdt'\int_V 2|U_1({\bf r})|\rho_0({\bf r}-{\bf
v}t',0)d^3r\right).
\label{wsol}
\ee

We further simplify our estimates and consider the step-like form for both functions, 
$U_1({\bf r})$ and $\rho_0({\bf r},t)$:
\be
U_1({\bf r}) =-\frac{\sigma_{in}v}{2V_1}\Theta(R_1-|{\bf r}|),\quad 
\rho_0({\bf r},t)=\frac{1}{V_2}\Theta(R_2-|{\bf r}-{\bf v}t|),\quad V_{1,2}=\frac{4}{3}\pi R_{1,2}^3.
\label{UT}
\ee

For the two limiting cases, $R_1 \gg R_2$ and $R_1 \ll R_2$,
the integral $\int 2|U_1| \rho_0 d^3r$
entering Eq. (\ref{wdot}) equals
\be
\frac{\sigma_{in}v}{V_1V_2}
\int \Theta(R_1-|{\bf r}|) \Theta(R_2-|{\bf r}-{\bf v}t'|)d^3r=
\frac{\sigma_{in}v}{V_{big}} \Theta(R_{big}-vt'),
\label{R1R2}
\ee
where
\be
R_{big}=max\left\{R_1,R_2\right\},\quad V_{big}=max\left\{V_1,V_2\right\},
\label{big}
\ee
and we have neglected the edge phenomena.

Now the integral over $t'$ in Eq. (\ref{wsol}) can be done trivially that
gives for the survival pro\-ba\-bi\-li\-ty:
\be
w_{surv}=\exp\left(-\frac{\sigma_{in}R_{big}}{V_{big}}\right),
\label{wlim}
\ee
which coincides with Eqs. (\ref{sur0}), (\ref{sur1}).

In general case, $R_1 \sim R_2$, the survival probability
can be presented as
\be
w_{surv}=\exp\left(-\frac{\sigma_{in}\bar R}{\bar V} 
F(R_1/R_2) \right),\quad {\bar R}=\sqrt{R_1^2+R_2^2},\quad 
{\bar V}=\frac43\pi{\bar R}^3,
\label{wgen}
\ee
with $F$ being a smooth function of $R_1/R_2$, $F(0)=F(\infty)=1$.

\begin{figure}[t]
\centerline{\epsfig{file=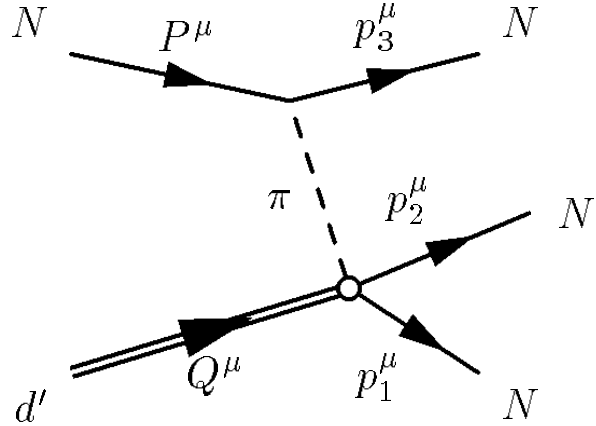,width=8cm}}
\caption{The diagram for the virtual pion exchange contribution to the spreading
width of the $d'$ resonance in the nuclear medium.}
\end{figure}

\section{Virtual pion exchange contribution to the spreading
width of the $d'$ resonance in nuclear medium}

In this section we estimate the spreading width of the $d'$ resonance in the nuclear
medium. Let us start from the virtual pion contribution described by the diagram in Fig.~1.

As shown in Ref. \cite{1}, at low energies there is only one Lorenz
invariant structure describing the $d'NN\pi$ vertex and, therefore,
the  amplitude of the decay $d' \to NN\pi$ can be written as
\be
\MM_{d' \to NN\pi} = \frac {f}{2m_N} 
\bar u_1C\gamma_5 (i\tau_2 \vec {\tau})\bar u^T_2 \vec {\pi},
\label{a1}
\ee
where $u_{1,2}$ are bispinors and $C = \gamma_2 \gamma_0$. The 
coupling constant $f$ can be expressed through the $NN\pi$ decay width 
$\Gamma=\Gamma_{pp\pi^-}+\Gamma_{nn\pi^+}+\Gamma_{np\pi^0}$ as
\be
f \approx\sqrt{
\frac{128\pi^2\sqrt{2}M\Gamma}{3\eta_0(M-2m_N-m_\pi)^2}\sqrt{\frac {m_N}{m_\pi}}}, 
\label{a2}
\ee
where $m_N$, $m_\pi$ and $M$ are the nucleon, pion and $d'$ masses, respectively;
$\eta_0$ being the enhancement factor due to the $NN$ FSI in this decay,
$\eta_0\approx 4 \div 5$. For $\Gamma \approx 0.5MeV$ (as deduced from the data on
the DCX reactions to discrete levels \cite{BCS}) $f \approx 14$.

The invariant matrix element of the process $d'N \to NNN$, when all
initial and outgoing particles are on mass shell, can be written in the form
\be
\MM^{(0)} = \sum_{ijk} \varepsilon_{ijk} \MM^{(0)}_{ijk} =
\frac{fg}{4m_N} \sum_{ijk} 
\bar u_iC\gamma_5 (\tau_2 \vec {\tau})\bar u_j^T
\frac{\varepsilon_{ijk}}{(P - p_k)^2 - m_\pi^2}\bar u_k \gamma_5 \vec{\tau}u,
\label{a3}
\ee
$$
\sum_{i=1}^3p_i^\mu=P^\mu+Q^\mu,
$$
where $g$ is the pseudoscalar $NN\pi$ coupling, $g^2/4\pi \approx 14.3$, 
$\varepsilon_{ijk}$ is the totally anti-symmetrical tensor, and indices $i,j,k = 1,2,3$
numerate the outgoing nucleons. As a result, the amplitude $\MM^{(0)}$ is totally
antisymmetric under permutation of the final nucleons. See also Fig.~1 where the
notations are explained.

The matrix element $\MM^{(0)}$ contains three different contributions, hence,
when squared, it produces three different diagonal terms
and three cross terms. After integration over the three-particle 
phase space all diagonal terms equally contribute to the cross
section. The same holds true for the cross terms. Hence, it is sufficient to
consider only one diagonal term squared ($e.g.$, $\{ijk\} = \{123\}$)
and one of the cross terms ($e.g.$, $\{ijk\}\times\{i'j'k'\}
= \{123\}\times\{321\}$).

The diagonal term $\{123\}$ squared and averaged/summed
over spins and isospins of the initial/final nucleons equals to:
\be
\langle |\MM^{(0)}_{123}|^2\rangle = 12\frac{(fg)^2}{m_N^2}(p_1p_2+m_N^2) 
\frac{(Pp_3-m_N^2)}{[(P-p_3)^2-m_\pi^2]^2}.
\ee
 
Integration over the phase space is essentially simplified if the
initial particles are at rest, {\it i.e.} $P^\mu=(m_N,0,0,0)$ and $Q^\mu=(M,0,0,0)$, which
means the limit $v\to 0$. In this limit $\sigma_{in}=\sigma_{d'N\to NNN}\propto 1/v$, 
so that the product $\sigma_{in}v$ remains constant. Thus, in this limit,
\be
\langle |\MM^{(0)}_{123}|^2\rangle = 24\frac{(fg)^2}{m_N} (E_3-m_N)\frac
{(M+m_N)^2+m_N^2-2(M+m_N)E_3}{(m_\pi^2-2m_N^2+2m_NE_3)^2},
\ee
{\it i.e.} $\langle |\MM^{(0)}_{123}|^2\rangle$ depends only on the energy of one of the nucleons, $E_3$,
which ranges between $m_N$ and 
\be
E_{max}=\frac{(M+m_N)^2-3m_N^2}{2(M+m_N)}.
\ee

Let us consider now the FSI between the nucleons in the relative $s$--state. 
Since there are three identical nucleons in the exit channel, only two of them can 
be in the relative s-wave with the third one being in the relative p-wave.
It is easy to see
that the matrix element $\MM_{ijk}^{(0)}$ (nucleon number {\it k} being in the
$NN\pi$ vertex) is proportional to the three-momentum $|{\bf p}_k|$,
therefore this nucleon is in $p$--wave in the rest frame of the initial
$d'$ and $N$ and, hence, in $p$--wave relative to the nucleons {\it i}
and {\it j} in the $d'NN\pi$ vertex (see Fig. 1). At the same time 
nucleons from the $d'NN\pi$ vertex are in relative $s$--wave \cite{1}. 
It means that only FSI between the nucleons in the $NN\pi$ vertex is 
important. It can be taken into account by multiplying 
$\MM_{ijk}^{(0)}$ by $\psi_{\bf  q}(0)$, where $\psi_{\bf q}(\bf r)$ is
the continuum $NN$ wave function (${\bf q}\approx ({\bf p}_i -{\bf p}_j)/2$)
containing the $s$--wave $NN$ scattering amplitude, if the $d'NN\pi$ vertex
is assumed to be point-like:
\be
\MM_{ijk} \approx \MM^{(0)}_{ijk} \left (1 + \frac{R^{-1}}
{-a_s^{-1}-iq_{ij}}\right),
\label{a4}
\ee
where $R \approx 0.8fm$ \cite{1}, $a_s$ is the $^1S$ scattering length,
and ${\bf q}_{ij}$ is the three-momentum 
of either  nucleon, {\it i} or {\it j}, in their centre-of-mass frame,
\be
q_{ij} = \sqrt {\frac {1}{4}[(M+m_N)^2+m_N^2-2E_k(M+m_N)] - m_N^2}.
\label{a5}
\ee

The Coulomb effects, which are important for very small invariant masses 
in the $pp$-subsystem, can be neglected in the integrated cross section.

With the FSI taken into account, the diagonal term 
has to be replaced by
\be
\langle|\MM^{(0)}_{123}|^2\rangle\to\langle|\MM_{123}|^2\rangle=\langle|\MM^{(0)}_{123}|^2\rangle 
\left |1 + \frac{R^{-1}}{-a_s^{-1}-iq_{ij}}\right|^2.
\ee

Finally, the summary contribution of all diagonal terms to the differential
cross section is
\be
d\sigma_{diag} = \frac {3}{(2\pi)^5}\frac {1}{3!}\frac {1}{4m_NMv}\langle|\MM_{123}|^2\rangle
\frac{d^3p_1}{2E_1} \frac{d^3p_2}{2E_2} \frac{d^3p_3}{2E_3}
\delta^4(Q+P-p_1-p_2-p_3),
\label{sdiag}
\ee 
where 3! accounts for the three identical particles in the final state.

As already mentioned, in the limit $v \to 0$ the integration over the 
phase space is simplified since there is no angular dependence, and
\be
\int\langle|\MM_{123}|^2\rangle
\frac{d^3p_1}{2E_1} \frac{d^3p_2}{2E_2} \frac{d^3p_3}{2E_3}
\delta^4(Q+P-p_1-p_2-p_3) = \pi^2 \int \langle|\MM_{123}|^2\rangle dE_2dE_3.
\label{sdiag2}
\ee

Now let us consider contributions of the cross terms ($e.g.$, $\{123\}\times\{321\}$):
\begin{eqnarray}
-\langle\MM^{(0)}_{123}\MM^{(0)*}_{321}+
\MM^{(0)}_{321}\MM^{(0)*}_{123}\rangle=-3\frac{(fg)^2}{m^2}\hspace*{3cm}\\
\times
\frac{(p_2p_3)(Pp_1-m_N^2)+(p_1p_2)(Pp_3-m_N^2)+m_N^2(P\sum_{i}p_i)-(p_1p_3)(Pp_2+m_N^2)-m_N^4}
{[(P-p_3)^2-m_\pi^2][(P-p_1)^2-m_\pi^2]}.\nonumber
\end{eqnarray}

As $v\to 0$ this expression depends only on energies of the outgoing nucleons,
\begin{eqnarray}
-\langle\MM^{(0)}_{123}M^{(0)*}_{321}+\MM^{(0)}_{321}M^{(0)*}_{123}\rangle=-3\frac{(fg)^2}{m}
\label{qwe}\hspace*{3cm}\\
\times\left\{\vphantom{\frac12}(E_1-m_N)[E_3(M+m_N)-m_N^2]+(E_3-m_N)[(E_3-m_N)[E_1(M+m_N)-m_N^2]\right.\nonumber\\
-M\left.\left[\frac{1}{2}(M^2-4m_N^2)-(M-E_1-E_3)(M+m_N)\right]\right\}
\frac{1}{[(P-p_3)^2-m_\pi^2][(P-p_1)^2-m_\pi^2]},\nonumber
\end{eqnarray}
and the FSI can be taken into account in the way described above, so that each 
$\MM^{(0)}_{ijk}$ has to be replaced by $\MM_{ijk}$, as in Eq. (\ref{a4}).

The integration over the three-particle phase space is similar to the integration
performed in Eq. (\ref{sdiag2}), so that for the total inelastic cross section one finds
\be
\sigma_{in}=\frac{4.5}{v}~mb,
\label{sin}
\ee
where the contribution of the cross terms is positive and does not exceed 15\% when $v\to
0$. The product $\sigma_{in}v$ is a smooth
function of the energy over the threshold, $\varepsilon$, so that, $e.g.$, for
$\varepsilon=20MeV$ (which corresponds to the kinetic energy of the nucleon 
of about $30MeV$ in the $d'$
rest frame) the contribution of the diagonal terms is only 10\%
larger than for $v\to 0$.

To proceed further we consider the $d'$ propagation in the nuclear medium, which we assume
infinite, for simplicity. Then, similarly to (\ref{twos}), one has for the spreading width:
\be
\Gamma_s=2|{\rm Im}U_A|.
\label{x1}
\ee
Here $U_A$ is the summary effective potential created by nucleons in the nuclear matter,
\be
U_A=\sum_{i=1}^{N_1}U_i\approx \rho_AV_1(U_0+iU_1),
\ee
where $\rho_A=\frac12m_\pi^3$ is the density of the nuclear matter, $N_1=\rho_A V_1$ being
the number of nucleons in the interaction region of the volume $V_1$. Using Eqs.
(\ref{U1}) and (\ref{x1}) one easily finds:
\be
\Gamma_s=2\rho_AV_1|U_1|=\rho_A\sigma_{in}v,
\ee
or, if the Pauli blocking factor, $\eta\approx 2.2$, is taken into 
account\ftnote{1}{To estimate Pauli blocking in the reaction $d'N \to NNN$ we simply limit the phase
space for the final nucleons by the constraint $p_N > P_F$, where $P_F$ is the Fermi
momentum.}, then  
\be
\Gamma_s=\rho_A(\sigma_{in}/\eta)v.
\label{kuku}
\ee

With the help of the result (\ref{sin}) one can estimate the spreading width to be about $7MeV$.
Notice, however, that the pion exchange (the diagram in Fig. 1) contributes only $30\div 40$\%
of the observed spreading width in nuclei, {\it i.e.}, there must be an additional mechanism
for the reaction $d'N\to NNN$, the total spreading width being
\be
\Gamma_s\approx 10\div 20 MeV.
\label{Gs}
\ee
Such a mechanism could be $\sigma$-exchange as discussed in Ref. \cite{KF}.         

Now we can check how well the approximation of the infinite nuclear medium works. To this
end we have to ensure that the typical free-path length of the $d'$ due to collision
damping, $L_{d'}\sim v/\Gamma_s$, does not exceed the radius of the nucleus $R_A$
$(\frac43\pi R_A^3\rho_A\equiv N_A)$. For DCX to discrete levels this condition is fulfilled
since for $\Gamma_s$ given by (\ref{Gs}) and $v\sim 1/10$, which corresponds to the 
energy $\varepsilon\approx 3MeV$ above the threshold, both lengths appear to be of the 
same order of magnitude, and $L_{d'}\lesssim R_A$.

Besides we can perform {\it a posteriori} check of the validity of the approximations made
in Section II. Namely, we can justify Born approximation used in 
Eq. (\ref{Im}):\ftnote{2}{Let us remind the reader that Born approximation is valid for
$|U_{0,1}|\ll 1/(\mu R_1^2)$.}
\be
\mu R_1^2|U_{0,1}|\approx\frac{\eta}{4\pi}\frac{m_N\Gamma_s}{\rho_AR_1}\approx\frac13,
\ee
and the one used in Eq. (\ref{opth})
when the elastic part of the total cross section was neglected.
Indeed, for the scattering off the step-like potential (\ref{U}) 
one has 
\be
\sigma_{el}=\frac{16}{9}\pi \mu^2R_1^6(U_0^2+U_1^2),\quad 
\sigma_{in}=\frac{8}{3}\pi R_1^3 U_1\frac{1}{v},
\label{sig}
\ee
with the ratio
\be
\frac{\sigma_{el}}{\sigma_{in}} = \frac{2}{3}\mu^2R_1^3
\frac{U_0^2+U_1^2}{|U_1|}v\approx\frac{\eta}{9\pi}
\frac{m_N^2 \Gamma_s}{\rho_A}v\approx \frac{1}{10},
\label{ratio}
\ee
where we put $|U_0|\sim |U_1|$ and substituted the velocity $v\sim 1/10$.

\section{Numerical estimates}

In this section we present the results of numerical calculations using Eqs.
(\ref{wgen}), (\ref{kuku}), and the estimate (\ref{Gs}). Thus for the survival
probability one, finally, has
\be
w_{surv}=\exp\left(-\frac{\eta{\bar R}\Gamma_s}{\rho_A{\bar V}v}\right),
\label{ws}
\ee
where we put the function $F(R_1/R_2)$ equal to unity everywhere for the sake of
simplicity.

In Fig. 2 we depict the survival probability (\ref{ws}) as function of the
excess energy of the initial pion over the threshold for $d'$ production 
using for the spreading width $\Gamma_s=20MeV$ and for the
radius $\bar R$ the values $1.4fm$ and $1.6fm$, respectively. For the 
excess energy of $3MeV$ above the threshold we find that only about
10\% of all created resonances survive, so that the suppression factor is of
order ten. Finally, in Fig. 3, we give the \lq\lq bare" total cross section of the $d'$
production in the reaction $\pi^3He\to d'N$ as well as the same cross section multiplied by
the factor of the survival probability. We find our theoretical predictions to comply 
reasonably well with the experimental data given in Ref. \cite{JG3} (dots with error bars
in Fig. 3), where already
the effects of collision damping had been discussed briefly.

\begin{figure}[t]
\centerline{\epsfig{file=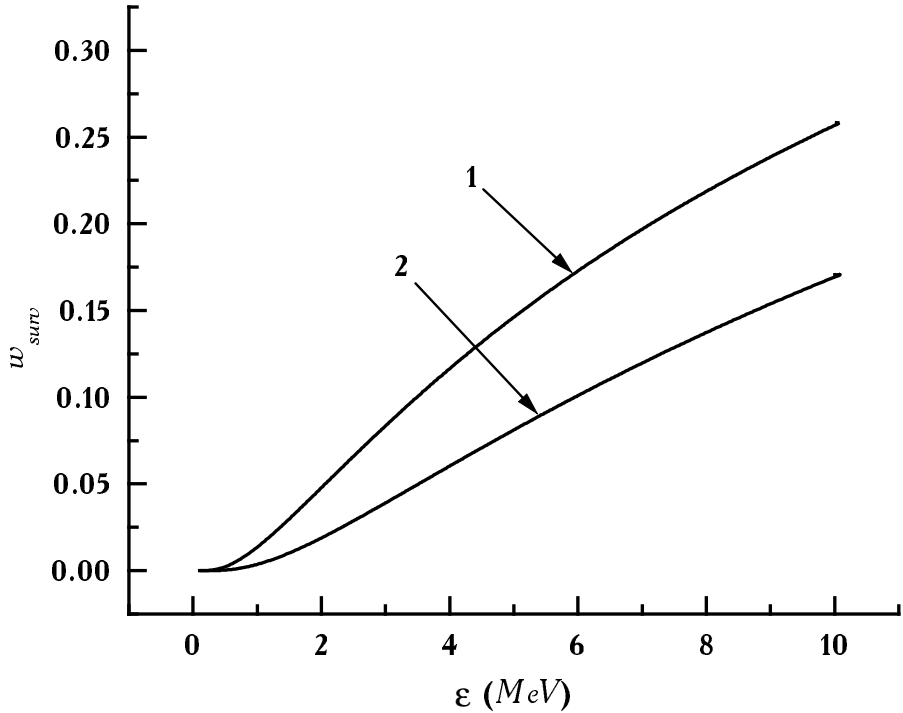,width=15cm}}
\caption{Survival probability $w_{surv}$ versus the excess of the pion kinetic energy 
over the threshold (in $MeV$) for $\Gamma_s=20MeV$ and $\bar{R}=1.6fm$ (the curve 1) and $\bar{R}=1.4fm$ 
(the curve 2).}
\end{figure}

\begin{figure}[t]
\centerline{\epsfig{file=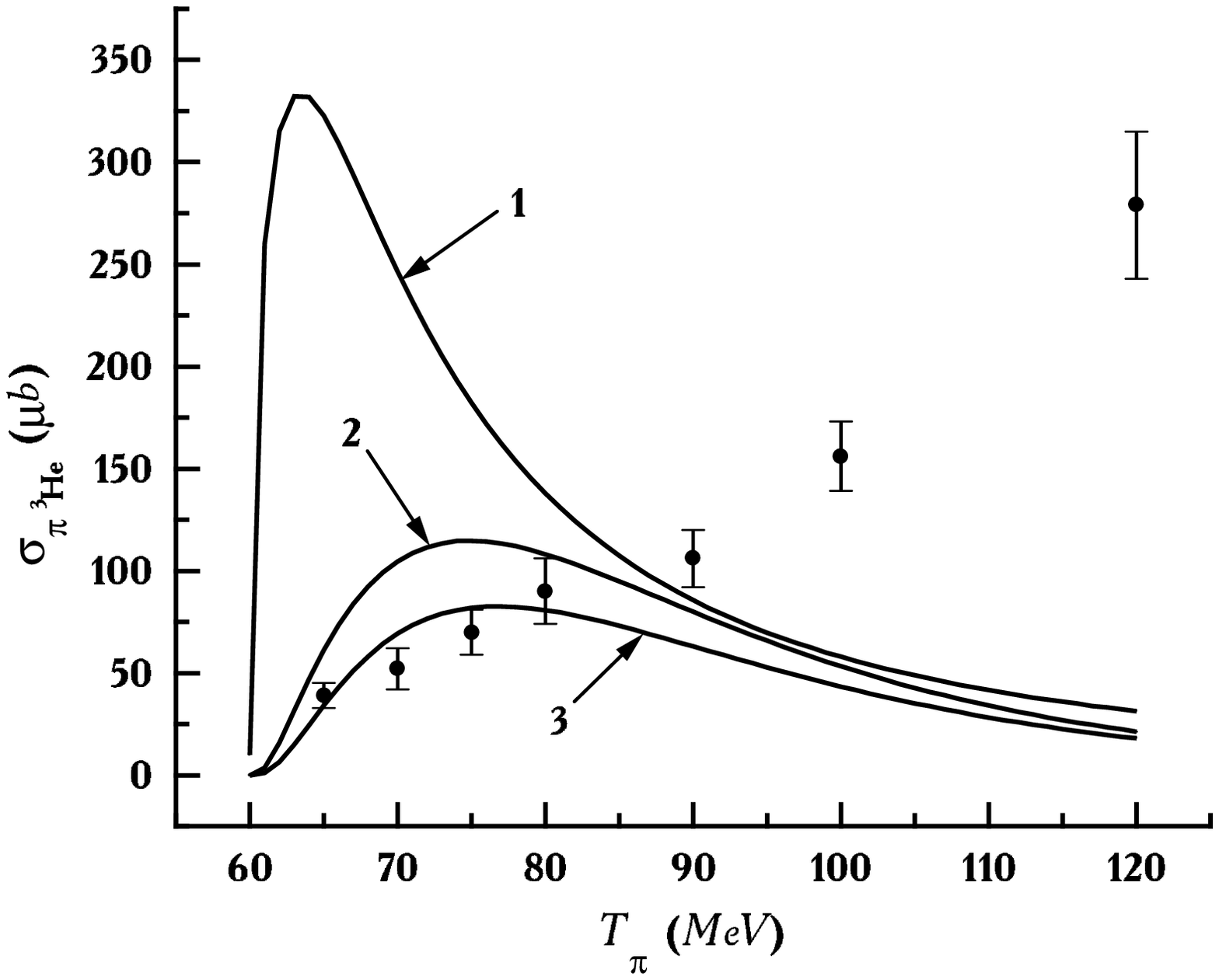,width=15cm}}
\caption{The cross section for the $d'$ production in the reaction 
$\pi^3He\to d'N$ (in $\mu barn$) versus the kinetic energy of the pion $T_{\pi}$ 
(in $MeV$). The curve 1 gives the \lq\lq bare" result, without the damping
effect taken into account \cite{CSWZ}; the curves 2 and 3 represent the same cross section
multiplied by the survival probability, Eq. (\ref{ws}), for $\Gamma_s=20MeV$ 
and ${\bar R}$ taking the values $1.6fm$ (the curve 2) and $1.4fm$ (the curve 3).
Experimental data \cite{JG3} are shown by the dots with error bars.}
\end{figure}

\section{Conclusions and outlook}

Using an optical potential approach we have shown that collision damping strongly 
decreases the survival probability of $d'$ in presence of a nuclear medium. Even
for $^3He$ it reduces the $d'$ production cross section near threshold by an 
order of magnitude, thereby leading to a reasonable agreement with the data.
Another way to approach the problem of collision damping of the $d'$
resonance near its production threshold would be a full coupled-channel
treatment. This requires the consideration of (at least) three channels:
$\pi^3He$, $d'N$ and $NNN$ (one could even neglect the vacuum width
of the $d'$ considering it as a stable particle). However, solving
the coupled-channel problem requires a rather accurate knowledge of
many reaction amplitudes (with $J^P = 1/2^-$) involved in this problem.
In principle, this would allow us to find the correction to the $d'$
production amplitude by summing up all possible contributions to
the amplitude $\pi^3He \to d'N$ including the route $\pi^3He\to NNN\to d'N$.
 
Let us note that such effects should equally
renormalize the amplitude of the $d'$ formation in heavier nuclei.
However, it is just this amplitude (deduced from DCX transitions
to discrete levels \cite{BCS}) that was taken as an input to predict the
$d'$ production off $^3He$ \cite{CSWZ}. The difference between these phenomena 
is, therefore, due to different details of the $d'$ propagation
following its initial production by the incoming pion.
In this respect it appears noteworthy to compare with the situation of
$\Lambda$ and $\Sigma$ production in the reaction $pp \to KNY$ near threshold, where
the observed \cite{sigma} surprisingly small $\Sigma$ production cross section is 
interpreted as being due to $pp \to K^+ p\Sigma^0 \to K^+ p\Lambda$, {\it i.e.},
due to a strong FSI between $p$ and $\Sigma^0$, which transfers $\Sigma^0$ immediately
into the energetically much more likely $\Lambda$ --- a situation very similar to 
that of $d'N \to NNN$ discussed above for the DCX on Helium isotopes.

\begin{acknowledgments}
One of the authors (A.V.N.) would like to thank the
staff of the Centro de F\'\i sica das Interac\c c\~oes Fundamentais (CFIF-IST) for cordial
hospitality during his stay in Lisbon. This work is supported by RFFI grants
00-02-17836, 00-15-96786, 02-02-06477 and 00-15-96562, INTAS grants 
OPEN 2000-110 and YSF 2002-49, 
BMBF (06 TU 987), DAAD (Leonhard-Euler Program), and DFG (Graduiertenkolleg).
\end{acknowledgments}

\end{document}